\documentclass[apj]{emulateapj}

\def\kms{{\rm\,km\,s^{-1}}}

\def\msun{{\rm\,M_\odot}}

\def\s{\ifmmode \widetilde \else \~\fi}
\def\={\overline}

\def\spose#1{\hbox to 0pt{#1\hss}}

\def\lta{\mathrel{\spose{\lower 3pt\hbox{$\mathchar"218$}}
     \raise 2.0pt\hbox{$\mathchar"13C$}}}
\def\gta{\mathrel{\spose{\lower 3pt\hbox{$\mathchar"218$}}
     \raise 2.0pt\hbox{$\mathchar"13E$}}}
\def\Dt{\spose{\raise 1.5ex\hbox{\hskip3pt$\mathchar"201$}}}    
\def\dt{\spose{\raise 1.0ex\hbox{\hskip2pt$\mathchar"201$}}}    

\def\dotsfill{\leaders\hbox to 1em{\hss.\hss}\hfill}

\long\def\symbolfootnote[#1]#2{\begingroup\def\thefootnote{\fnsymbol{footnote}}\footnote[#1]{#2}\endgroup}

\begin{document}

\title{What Sets the Sizes of the Faintest Galaxies?}

\author{Crystal M. Brasseur$^{1,3}$,  Nicolas F. Martin$^1$, Andrea V. Macci\`o$^1$, Hans-Walter Rix$^1$, Xi Kang$^2$ }
\affil{$^1$Max-Planck-Institut f\"{u}r Astronomie, 17 K\"{o}nigstuhl 69117 Heidelberg, Germany\\
$^2$The Purple Mountain Observatory, 2 West Beijing Road, Nanjing 210008\\
$^2$deceased}
\email{martin,maccio,rix@mpia.de}

\begin{abstract}

We provide a comprehensive description and offer an explanation for the sizes of the faintest known galaxies in the universe, the dwarf spheroidal (dSph) satellites of the Milky Way and Andromeda. After compiling a consistent data set of half-light radii ($r_{1/2}$) and luminosities, we describe the size-luminosity relation of dSphs by a log-normal distribution in $r_{1/2}$ with a mean size that varies as a function of luminosity. Accounting for modest number statistics, measurement uncertainties and surface brightness limitations, we find that the size-luminosity relations of the Milky Way and Andromeda dSph populations are statistically indistinguishable, and also very similar: their mean sizes at a given stellar luminosity differ by no more than 30\%. In addition, we find that the mean size, slope and scatter of this log-normal size description of Local Group dSphs matches onto the relation of more massive low-concentration galaxies. This suggests that the stellar sizes of dSphs are ultimately related to their overall initial baryonic angular momentum. To test this hypothesis we perform a series of high resolution N-body simulations that we couple with a semi-analytic model of galaxy formation. These predict the same mean size and slope as observed in dSph satellites. At the same time, these models predict that the size-luminosity distributions for satellite galaxies around similar host-halos must be similar providing a natural explanation as to why the size distributions of Milky Way and Andromeda satellites are similar. Although strong rotation is currently not observed in dSphs, this may well be consistent with our angular-momentum-based explanation for their sizes if the disks of these galaxies have become sufficiently stirred through tidal interaction.

\end{abstract}

\section{Introduction}
\label{sec:intro}

\symbolfootnote[0]{$^\star$deceased}
Galaxies with stellar masses of $10^8-10^{11.5}\msun$ have a well defined relation between their stellar mass and the characteristic radius of their stellar body. This relation is one aspect of well-established global galaxy-parameter relations, such as the fundamental plane or the Tully-Fisher relation \citep{kormendy77,tully77,djorgovski87,dressler87}. For disk galaxies, the size of the stellar body is generally related to the angular momentum created by torques produced during the hierarchical formation of a galaxy and the infall of material towards the center of the gravitational potential well. In-falling gas dissipates and cools, but can only collapse until angular momentum prevents further central concentration \citep{fall80,mo98,shen03}.  This  implies a size scaling with stellar mass, with the latter relating to the halo size, the baryonic fraction and the halo spin parameter.

This simple angular momentum description is modified by various mechanisms of galaxy formation and evolution: major merging for massive early-type galaxies, questions relating to angular momentum conservation, bar instabilities (which can redistribute internal angular momentum: \citealt{athanassoula03,kormendy89,vandenbosch98}), as well as star formation efficiency. Despite these complicating effects, however, a simple angular-momentum-based formulation has been shown to provide a plausible explanation for the sizes of most galaxies at both low (e.g., \citealt{shen03}) and high redshifts \citep{somerville08}.  
At the low mass end of the galaxy scale, however, the size distribution of dwarf spheroidal galaxies (dSphs) which contain at most a few million stars is poorly determined empirically and without a cogent explanation. 

The sizes of Local Group dSphs came into focus when \cite{mcconnachie06b} pointed out that bright dwarf galaxies of M31 were larger by a factor of 2-3 compared to their Milky Way counterparts of the same luminosity.  As we will show later on in the paper, such an apparent size difference is difficult to reproduce in a cosmological framework if the stellar sizes reflect halo sizes or halo angular momenta: N-body simulations fail to explain a size discrepancy in sub-halo populations through differences in either host halo collapse-time or host halo masses. \cite{penarrubia08b} investigated whether a varied amount of tidal stripping between the Andromeda and Milky Way satellites could produce the observed size difference. They found that tidal effects could cause the reported difference in the characteristic sizes, but that the tides would also simultaneously lower their surface brightness. On the contrary, M31 dwarf galaxies appear to be both larger and lower surface brightness than Milky Way dwarf galaxies, thereby challenging this scenario. One caveat of these simulations is that they do not include a baryonic disk. Later on, \citet{penarrubia10} have shown that including such a disk can have a significant impact on the evolution of satellite dwarf galaxies. As a consequence, a difference in the stellar mass of the Milky Way and M31 disks could be driving some of the apparent differences between their satellite populations. This analysis does not, however, explicitly study the possibility of induced changes in size of the simulated dwarf galaxies.

There are several practical difficulties in determining the sizes of faint, low surface brightness dwarf galaxies ($M_V>-16$). These galaxies can only be studied in detail within the Local Group, where they can be resolved into their constituent stars: a single dwarf galaxy can contain as little as a few tens of stars down to current survey limits and determining their half-light radius, $r_{1/2}$, requires a suitable mathematical approach (e.g. \citealt{martin08b,brasseur11a}). In addition, survey detection limits and, therefore, the incompleteness in dwarf galaxy searches, play an important role in how we observe and build up their size-luminosity relation. If detection limits are not taken into account, one will invariably find that fainter galaxies tend to be smaller if large, faint galaxies (which have a lower surface brightness) fall below the detectability threshold.

Since the work of \cite{mcconnachie06b}, the number of known Local Group satellites has strongly increased, (e.g. the number of known Andromeda dSphs has more than tripled).  Additionally, since this time, satellites have been discovered more than a factor of 10 fainter in luminosity, and we have increased our understanding of the surface brightness limits of both Milky Way and Andromeda surveys. 

In this paper we set out to statistically quantify for the first time the global size-luminosity distribution of dSph satellites around the Milky Way and Andromeda, using a mathematical description which accounts for the detection limits and low-number statistics. We then proceed to address the following questions: Are there differences in the mean sizes of Milky Way and Andromeda satellite, as previously pointed out? How does the size-luminosity relation at the extreme low-mass end of galaxies fit into a cosmological context?  Does an interpretation of angular momentum setting the characteristic size of galaxies apply to the very low mass end, and does this fit with current observations?

In section \ref{sec:size_description}, we compile a state-of-the-art data set of the sizes and magnitudes of low luminosity galaxies around the Milky Way and Andromeda. We then develop and apply a technique to describe the size--magnitude relation for galaxies around Andromeda and the Milky Way in section \ref{sec:ML}. We explore any possible size differences between the sub-populations before comparing the  $r_{1/2}-M_V$ relation found here (mean size, slope, scatter) to that of more massive galaxies (\citealt{shen03}; section \ref{sec:sizesCompare}). We put this in the context of cosmological expectations in section \ref{sec:Theoretical} and finally conclude on the implications for explaining the sizes of the smallest galaxies in section \ref{sec:discussion}.

\section{The Sample}
\label{sec:size_description}

In this study we are interested in only the faintest satellite galaxies and, as such, the Milky Way and Andromeda satellite systems are the only two laboratories available for studying in detail the size-luminosity relation of dSphs. Our sample therefore consists of all detected satellites of the Milky Way and Andromeda within a stellar mass range equivalent to $-14<M_V <-6$. We note that throughout this paper we use absolute magnitude and stellar mass almost interchangeably, since the stellar mass-to-light ratios of these dwarf galaxies have little scatter (they are mainly old, metal-poor stellar populations) and vary smoothly with stellar mass.



As much as we can, we strive to construct a sample of homogeneously measured sizes and luminosities of all Milky Way and Andromeda dSphs which we correct for the effects of surface brightness limitations.

\subsection{The Sizes and Luminosities of Milky Way and Andromeda Satellite Dwarf Galaxies}

One of the difficulties of putting together a homogeneous sample of Local Group dwarf galaxy properties comes from the strongly heterogeneous values that can be found in the literature, measured using different data sets, and different techniques. We have therefore tried as much as possible to restrict ourselves to a small set of comprehensive studies. Given these considerations, we rely namely on \citet{irwin95} and \citet{mateo98} for the bright Milky Way dwarf galaxies, and \citet{martin08b} for the recent Milky Way dwarf galaxy discoveries. For bright Andromeda dwarf galaxies, we rely on \citet{mcconnachie06b}, updated with measurements from dwarf galaxy discovery papers from the Pan-Andromeda Archaeological Survey(PAndAS; \citealt{martin06b,martin09,ibata07,mcconnachie08,richardson11}), a photometric survey of the M31/M33 group conducted with the MegaCam wide-field camera on the Canada-France-Hawaii Telescope \citep{mcconnachie09}. The PAndAS data themselves have been partially superseded by deeper photometric follow-up data sets that we have obtained for a large fraction of these satellites (\citealt{collins10,brasseur11a,brasseur11b}). Detailed references are listed in Tables~\ref{mw_table} and~\ref{and_table}.

For dwarf galaxies with $M_V<-8$, size and luminosity measurements are quite robust and depend little on the method used to obtain them. Dwarf galaxies with $M_V>-8$, on the other hand, usually contain only a few tens or hundred stars down to the observational limit from which one has to reliably measure a size and a luminosity. Some of the difficulties inherent to this problem are described in detail in \citet{martin08b} which also presents a maximum likelihood algorithm to measure the structural parameters of a dwarf galaxy. We rely as much as we can on publications that have used this technique to derive the size of the galaxies in our sample. For details on its applications, the reader is referred to \citet{martin08b} for the study of faint Milky Way satellites and \citet{brasseur11a} for an application to faint Andromeda satellites.

An additional difficulty comes from the use of different effective radii in the literature. We have homogenized the data set by using, whenever possible, half-light radii measured along the major axis of the systems, and with the assumption that the radial density (or light) profile of the galaxy is exponential. When only the exponential scale radius is available, it has been converted to $r_{1/2}$ by multiplying by 1.68. Values that correspond to the geometric mean of the minor and major axes half-light radii have been converted using the ellipticities given in the corresponding study.

For estimating total stellar luminosity, \citet{martin08b} have shown that color-magnitude diagram shot-noise can become an issue but this is only the case for very faint systems, so not essential for Andromeda satellites, which all have $M_V<-6.0$

\subsection{(In)completeness Limits of Dwarf Galaxy Surveys}
\label{sec:sblimits}

All galaxy surveys suffer from some level of incompleteness; in the case of Local Group galaxies this arises from areal coverage, volume probed and the presence of  Milky Way foreground stars. For example, the Sloan Digital Sky Survey, SDSS, led to the discovery of many new dSphs, but, as the survey area samples only one quarter of the sky around the North Galactic Cap, many of the known Milky Way satellites are concentrated in this region of the sky and undiscovered dSphs are likely to lay in regions of the sky which have not been imaged with the depth of the SDSS.

Volume incompleteness issues do not, however, affect our final result if we assume the size-luminosity relation is not radially dependent. For our analysis we do, however, need to properly account for the luminosity and surface brightness limits of dSph searches around the Milky Way and Andromeda galaxies. 

\begin{figure}
\includegraphics[width=\hsize]{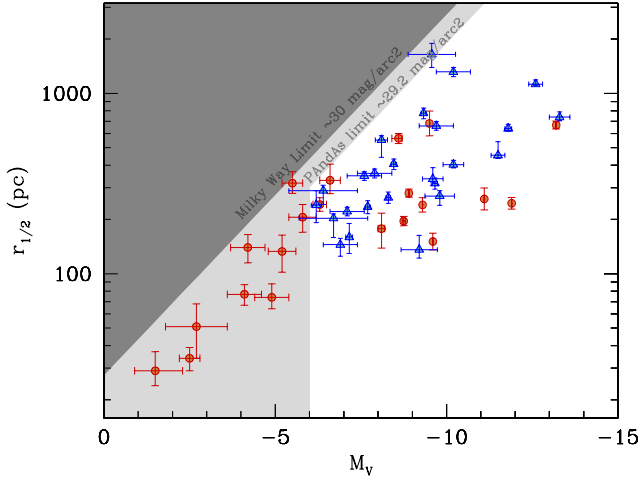}
\caption{Comparing the observed size distribution, $r_{1/2}$, of Andromeda low luminosity dwarf satellites galaxies (blue triangles) with those in the Milky Way (orange circles). The shaded regions correspond to the approximate observation limits of the Milky Way (dark grey) and Andromeda (light grey). }
\label{logMvr12}
\end{figure}

In the case of the Milky Way, \cite{koposov08} quantify the distance dependent surface brightness limits for discovering satellites in the SDSS. This broadly corresponds to a limit of $\sim 30\,\textrm{mag}/\textrm{arcsec}^2$. In the case of Andromeda, all the dwarf spheroidal galaxies lie at approximately the same distance from us, leading to a fixed surface brightness limit for discovering dwarfs. PAndAS has completed observations out to 150 kpc from the center of M31 to an average surface brightness of  $\sim 29.2\,\textrm{mag}/\textrm{arcsec}^2$ (the central surface brightness of And~XIX, the dwarf galaxy with the lowest surface brightness found in PAndAS), shown in light grey in Figure \ref{logMvr12}. Additionally, those galaxies with total magnitudes fainter that $M_V>-6$ would not be easily detected in PAndAS, since they would have only a few stars along there red giant branch, and would go undetected by the selection techniques currently employed to discover new dwarfs.

The PAndAS collaboration is currently in the process of accurately determining the completeness limits of their dwarf galaxy searches but this study can afford to rely only on these current rough estimates. We will show later that shifting the assumed detection limits for both Milky Way and Andromeda surveys to more conservative values does not change our results significantly.

\section{Determining the Stellar Size-Luminosity Distribution}
\label{sec:ML}

In this section, we outline our method for determining the size distribution of dSphs as a function of magnitude. Any practical approach must account for the uncertainties, modest number statistics, and the observational detection limits of each sample. We satisfy these requirements by adopting a parametrized functional form for the distribution and use a maximum likelihood algorithm to identify sets of parameters that make the observed data likely.

For our analysis we adopt a log-normal distribution for the sizes of dSphs. This is found to approximate the intrinsic size distribution of more massive low-concentration galaxies (\citealt{shen03}) and, as will be detailed in  section~\ref{sec:wholesample}, is also well motivated by theoretical considerations: for massive disk galaxies, the size of the stellar disk is generally related to the angular momentum, with the latter relating to the halo spin parameter. The log-normal distribution of $\lambda$ leads to a size distribution that is approximately log-normal.

We fit the Milky Way and Andromeda dSphs separately in [$\log r_{1/2}$, $M_V$]-space (seen in Figure \ref{logMvr12}), adopting the functional form from \citealt{shen03}:  a normal distribution in $\log r_{1/2}$ that has a slope, $S$, with magnitude and a dispersion, $\sigma_{\lg\,r}$, around this slope. With $\overline{\lg\,r}$ being the mean of $\log r_{1/2}$ at $M_V=-6.0$ and $\Delta M_V = M_V + 6.0$ being the magnitude offset from $M_V=-6.0$, we parametrize the mean of the distribution, $\overline{\lg\,r}^{\prime}$, as

\begin{equation}
\overline{\lg\,r} ^{\prime }=\overline{\lg\,r}+ S ~ \Delta M_V.
\end{equation}

A critical part of our analysis is accounting for the surface brightness limitations of each distribution. These limits are parametrized at the faint end as $M_V(\log r_{1/2})$, and are represented by the shaded areas in Figure \ref{logMvr12}. To properly account for these limitations, we integrate our likelihood function over the observable window in [$\log r_{1/2}$, $M_V$]-space in order to correctly recover the normalization constant, $A$, of our likelihood function.  Thus, $A$ is re-computed for each choice of $\overline{\lg\,r}$, $\sigma_{\lg\,r}$ and $S$ in the maximum likelihood grid, and we fix the normalization constant such that:

\begin{eqnarray}
n = \frac{A(\overline{\lg\,r},\sigma_{\lg\,r},S)}{\sigma_{\lg\,r} \sqrt{2\pi}}  \int_{\log r_{1/2}} \int^{M_V(x)}\nonumber\\
\times \exp\left[ - \frac{(x-\overline{\lg\,r}^{\prime })^2}{2 \sigma_{\lg\,r}^2} \right] dy dx,
\end{eqnarray}

\noindent where $n$ is the total number of dwarfs in the observable window.

We also want to account for the individual measurement errors of each dwarf. To do this, we parametrize the variance of our distribution, $\sigma_{\lg\,r}^{\prime 2}$, as the sum of the intrinsic variance of the log-normal distribution, $\sigma_{\lg\,r}^2$, and the error on each individual $\log r_{1/2}$ measurement, $\sigma_{\lg\,r,\,i}$:

\begin{equation}
\sigma_{\lg\,r} ^{\prime 2}=\sigma_{\lg\,r}^2+\sigma_{\lg\,r,i}^2,
\end{equation}

\noindent which reduces to $\sigma_{\lg\,r}^2$ in the limit of zero measurement errors. 
Therefore the likelihood of a set of $n$ dSphs, with the log of their half-light radius $x_i$, the corresponding uncertainties $\sigma_{\lg\,r,\,i}$, magnitudes $M_{V,i}$ such that $\Delta M_{V,i}=M_{V,i}+6.0$, can be expressed as:

\begin{eqnarray}
\mathcal{L} \left(\{x_i,\sigma_{\lg\,r,\,i},\Delta M_{V,i}\}|\overline{\lg\,r}, \sigma_{\lg\,r}, S\right)\hspace{3cm}\nonumber\\
=\prod_{i=1}^{n} \frac{A(\overline{\lg\,r},\sigma_{\lg\,r},S)}{\sigma_{\lg\,r}^{\prime}\sqrt{2\pi}} \exp\left[ - \frac{(x_i-\overline{\lg\,r}^{\prime})^2}{2 \sigma_{\lg\,r}^{\prime 2}}   \right]  \hspace{1.3cm}\nonumber\\
=\frac{2\pi^{-n/2} A^n(\overline{\lg\,r},\sigma_{\lg\,r},S)}{ \prod_{i=1}^{n} \sqrt{\sigma_{\lg\,r}^2+\sigma_{\lg\,r,i}^2}}\hspace{3.85cm}\nonumber\\
\times\exp\left[-\sum_{i=1}^{n}  \frac{(x_i-\overline{\lg\,r}- S~ \Delta M_{V,i})^2}{2(\sigma_{\lg\,r}^2+\sigma_{\lg\,r,i}^2)}   \right].\hspace{1.5cm}
\end{eqnarray}

In practice it is often more convenient to work with the logarithm of the likelihood function. Since the logarithm is a monotonically increasing function, the logarithm of a function achieves its maximum value at the same points as the function itself. Thus our likelihood function which we compute at each position within a fine grid in parameter space becomes:

\begin{eqnarray}
\ln \mathcal{L}\left(\{x_i,\sigma_{\lg\,r,\,i},\Delta M_{V,i}\}|\overline{\lg\,r}, \sigma_{\lg\,r}, S\right)\hspace{3.8cm}\nonumber\\
 = n \ln A(\overline{\lg\,r},\sigma_{\lg\,r},S) -\frac{n}{2}\ln(2\pi) - \sum_{i=1}^{n} \ln \left( \sqrt{\sigma_{\lg\,r}^2+\sigma_{\lg\,r,i}^2}  \right)\hspace{0.2cm}\nonumber\\
 -0.5\sum_{i=1}^{n}\frac{(x_i-\overline{\lg\,r}-S~\Delta M_{V,i})^2}{\sigma_{\lg\,r}^2+\sigma_{\lg\,r,i}^2}.\hspace{4.02cm}
\end{eqnarray}

From there, we determine the model that maximize the likelihood function by exploring a fixed grid of parameters $\overline{\lg\,r}$, $S$ and $\sigma_{\lg\,r,i}$ between 1.99 and 2.55, 0.01 and 0.60, and -0.4 and 0.0, respectively, and with step sizes of 0.01 in all cases.

\section{The dwarf galaxy size-luminosity relation of Milky Way and Andromeda satellites}
\label{sec:sizesCompare}

\begin{figure}
\includegraphics[width=\hsize]{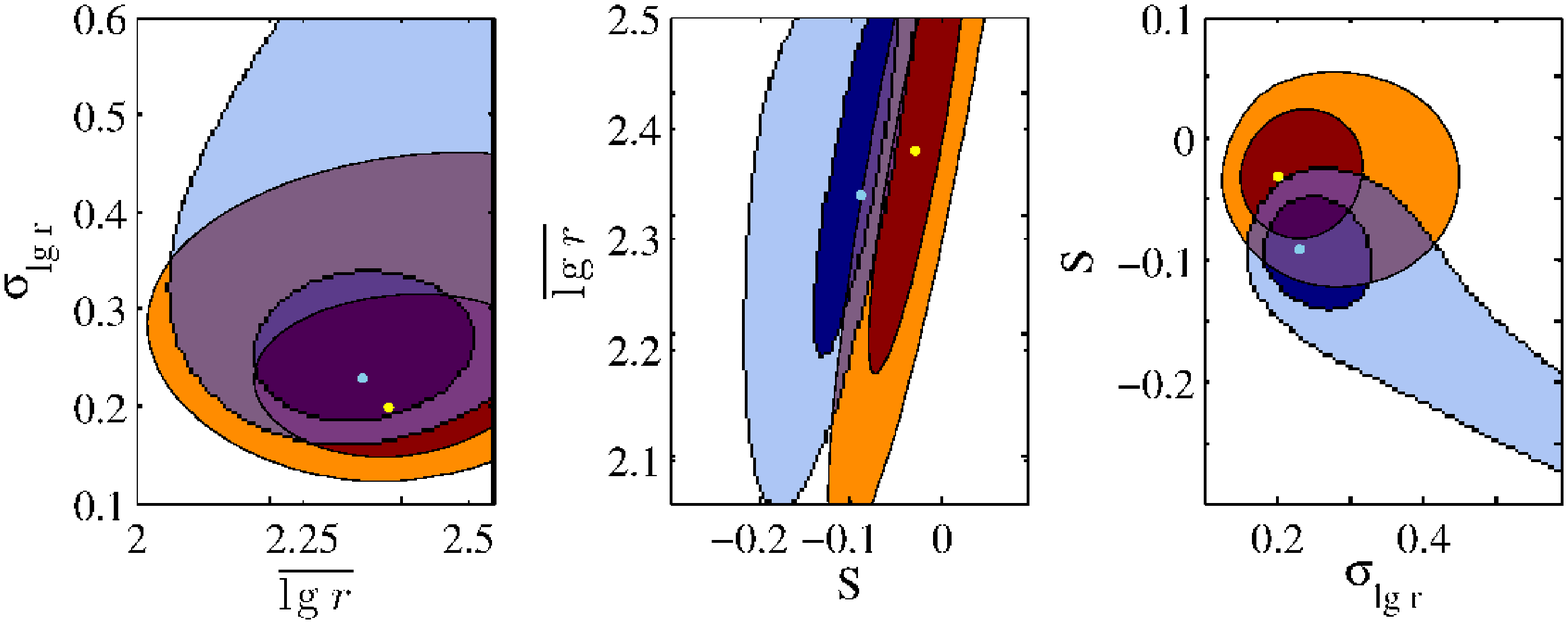}
\includegraphics[width=\hsize]{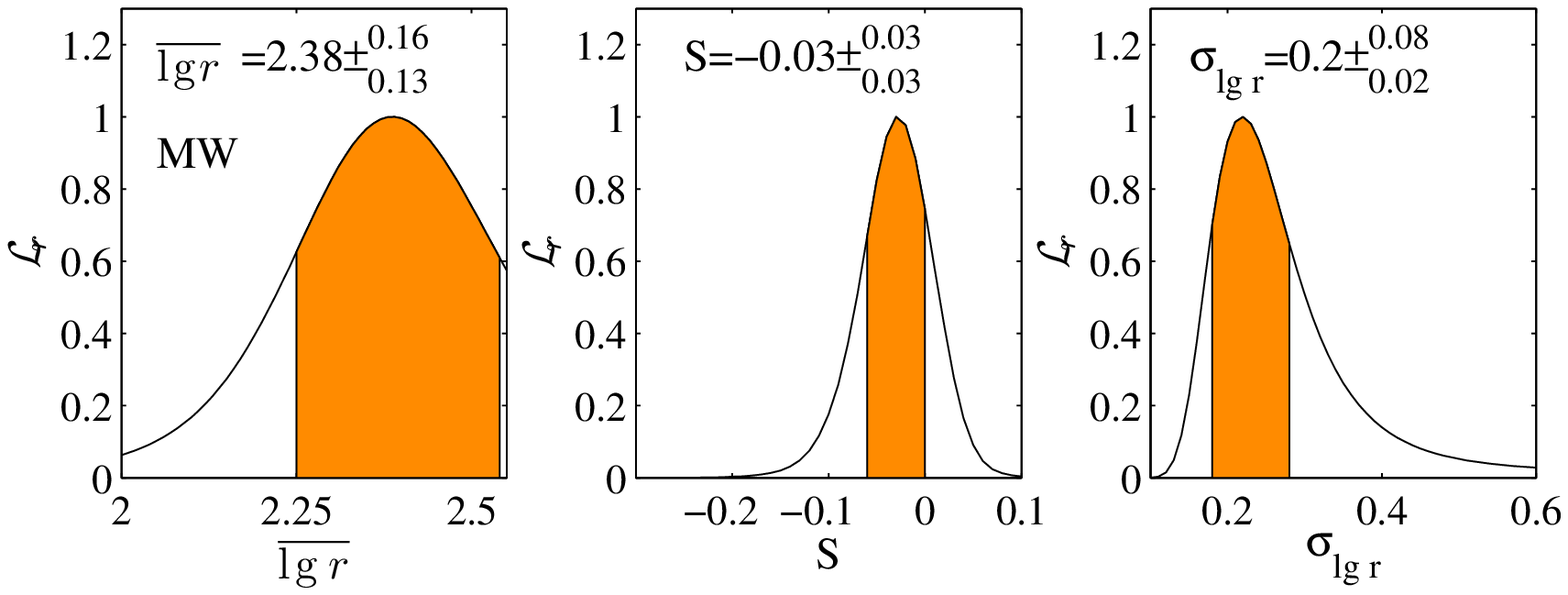}
\includegraphics[width=\hsize]{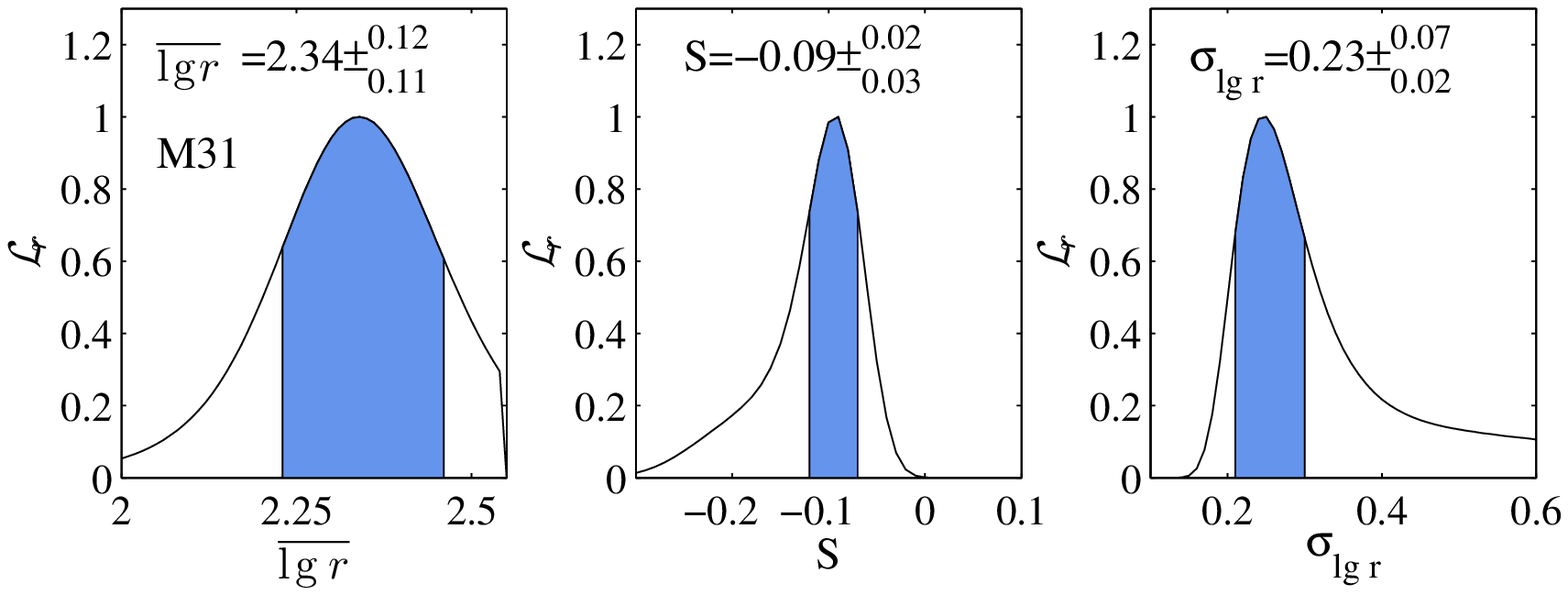}
\caption{Parameters of a log-normal size distribution for the Milky Way (orange) and Andromeda (blue) dSph populations, derived from our likelihood analysis. The distribution of dSphs is assumed to be log-normal in $r_{1/2}$, with a mean size at $M_V=-6$, $\overline{\lg\,r}$, an intrinsic dispersion, $\sigma_{\lg\,r}$, and a slope in magnitude, $S$.
{\bf Upper Row:} The two dimensional relative likelihoods of each pair of the three fitted parameters marginalized over the third model parameter. The contours correspond to the 1 and 2-$\sigma$ limits, derived from $\chi^2$ values with 2 degrees of freedom: $-2\, \ln \mathcal{L}_r=2.30$ and 6.17, respectively. The most likely values are shown by dots in yellow (Milky Way) and cyan (Andromeda). Although the maxima for the Milky Way and Andromeda occur at different points, there is significant overlap in the 1 and 2-sigma contours. Thus, our global analysis finds no significant difference between the Milky Way and Andromeda dSph populations.  
{\bf Middle Row:} The relative likelihood functions of $\overline{\lg\,r}$, $S$  and $\sigma_{\lg\,r}$, marginalized over the other two model parameters for Milky Way dSphs. The colored regions under the curves indicate the 1-$\sigma$ confidence intervals derived from $\chi^2$ values with 1 degree of freedom: $-2\, \ln \mathcal{L}_r=1$. Our computed values for each parameter are printed at the top of each window.
{\bf Lower Row:} Same as the middle row but for Andromeda dSphs.
}
\label{ML_3param}
\end{figure}

We show the results of our likelihood method in Figure \ref{ML_3param} for Milky Way (orange) and Andromeda (blue) dSphs.  For these results, we have restricted the maximum likelihood method to those satellites with $M_V<-6$ to ensure consistency between the Milky Way and Andromeda.

\subsection{Comparing the Milky Way and Andromeda satellite systems}

 In the top panel of Figure \ref{ML_3param} we overlay the resulting 1 and 2-$\sigma$ contours of the Milky Way (orange) and Andromeda (blue) size-luminosity parameters. Yellow and cyan dots correspond to the parameter values of the preferred models, i.e. the set of model parameters that maximizes the likelihood function for the Milky Way and Andromeda, respectively. The 1 and 2-$\sigma$ limits are derived from $\chi^2$ values with 2 degrees of freedom: $-2\,\ln \mathcal{L}_r=2.30$ and 6.17, respectively.  One can see from the top panel of Figure \ref{ML_3param}, that although the preferred models for the Milky Way and Andromeda are different, there is significant overlap of the 1 and 2-sigma contours.

 The lower two panels of Figure \ref{ML_3param} show the resulting relative likelihood, $\mathcal{L}_r= \mathcal{L}/ \mathcal{L}_{max}$, functions for $\overline{\lg\,r}$, $S$  and $\sigma_{\lg\,r}$, marginalized over the other two model parameters. The colored regions under these curves indicate the 1-$\sigma$ confidence intervals derived from $\chi^2$ values with 1 degree of freedom: $-2\, \ln \mathcal{L}_r=1$. The Milky Way and Andromeda values of $\overline{\lg\,r}$ ($2.38^{+0.16}_{-0.13}$ and $2.35^{+0.11}_{-0.14}$, respectively) and $\sigma_{\lg\,r}$ ($0.20^{+0.08}_{-0.02}$\,dex and $0.24^{+0.08}_{-0.02}$\,dex, respectively) are consistent within 1-$\sigma$. Only the determined $S$ values deviate slightly from each other, but only corresponds to a $1.7-\sigma$ deviation for this parameter, with $S=-0.03\pm0.03$ for the Milky Way and $S=-0.09^{+0.02}_{-0.04}$ for the Andromeda dSph population.

One caveat of our results is that our maximum likelihood method relies on our choice of surface brightness limits. As described in section \ref{sec:sblimits}, these limits are not precisely defined. To test how the choice of these limits affects our results, we have additionally run our maximum likelihood technique using more conservative surface brightness limits. We re-ran our algorithm for ten different choices of the surface brightness limits for the Milky Way and Andromeda, in steps of $0.2\,\textrm{mag}/\textrm{arcsec}^2$ towards brighter limits.  In each case, we again find significant overlap in the likelihood contours for Milky Way and Andromeda dSphs. We therefore conclude that for reasonable choices in the surface-brightness limits of the observations, our results remain unchanged.

In addition, this analysis is performed assuming that two of the largest Andromeda satellites, And XIX and XXVII are bound systems, while observations of their surroundings show that they are likely to be embedded in stellar streams \citep{mcconnachie08,richardson11}. This might hint that they are strongly deformed, and maybe inflated, by their interaction with Andromeda. Removing them from our sample would further diminish the small difference we measure between the size-luminosity relations of Milky Way and Andromeda dwarf galaxy populations.

To conclude, we find that when one accounts for completeness limits and the modest number of objects in the sample, there is no significant difference between the global size-luminosity relation of the Milky Way and Andromeda dSph populations.

\subsection{The size distribution of the combined sample}
\label{sec:wholesample}

\begin{figure}
\includegraphics[width=\hsize]{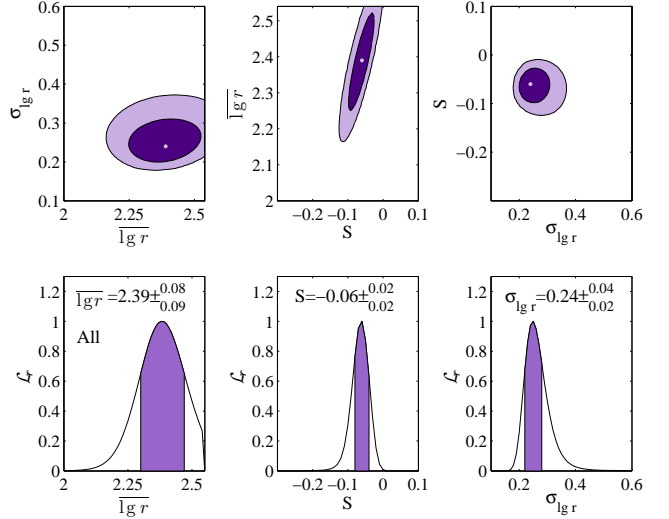}
\caption{Same as for Figure \ref{ML_3param} but with our maximum likelihood method applied jointly to all Milky Way and Andromeda dSphs.}
\label{ML_all}
\end{figure}

In the previous section, we find that the size-luminosity relations of Milky Way and Andromeda dSphs are statistically indistinguishable.  Building on this result, we assume a common size-luminosity relation for all dSphs by combining the Milky Way and Andromeda dSphs, and re-run our maximum likelihood method. In the lower two panels of Figure~\ref{ML_all} we show the resulting marginalized relative likelihood curves with the colored regions under these curves indicating the 1-$\sigma$ limits of each parameter. The favored values for the model parameters are $\overline{\lg\,r}=2.39^{+0.09}_{-0.09}$, $S=0.24^{+0.04}_{-0.02}$ and $\sigma_{\lg\,r}=0.06^{+0.02}_{-0.02}$\,dex. In the top panel of Figure \ref{ML_all} we plot the resulting 1 and 2-$\sigma$ contours for these parameters, and the light purple dots correspond to the most likely model.

\subsection{ A comparison with the sizes of more luminous galaxies}

Leaving the boundaries of the Local Group, one can wonder how the size-luminosity relation of dSphs compares to that of more massive galaxies. From a sample of 140,000 galaxies in the SDSS, \citet{shen03} study the size distribution of galaxies and its dependence on luminosity. They use a S\'ersic index of $n= 2.5$ to separate their sample into high-concentration, early-type ($n >2.5$) and low-concentration, late-type ($n <2.5$) galaxies.

The characteristic sizes of massive disk galaxies are generally related to the angular momentum of the system which is built up through torques during the process of hierarchical formation of a galaxy. Gas infalling into the gravitational potential well dissipates and cools, but is prevented from further collapse by the angular momentum of the system \citep{fall80,mo98,shen03}. Therefore the gas, and consequently the stars, will retain a size scaling with the initial angular momentum of the system. Treating dark halos as singular isothermal spheres, and neglecting the gravitation effects of the disks, \cite{mo98} derive the stellar disk scale length, $R_d$, to be:

\begin{equation}
R_d=\frac{1}{\sqrt{2}}\left(\frac{j_d}{m_d}\right) \lambda r_{200} f_r(\lambda, c_\mathrm{vir}, m_d, j_d), 
\label{Rd}
\end{equation}

\noindent where $m_d$ and $j_d$ are the fractions of mass and angular momentum of the halo that are in the disk, $\lambda$ is the halo spin parameter, $r_{200}$ is the radius within which the mean density is 200 times the critical density, and $f_r$ is a factor that depends on the halo profile and the action of the disk:

\begin{eqnarray}
f_r & = & \frac{\lambda(j_d/m_d)}{0.1}^{-0.06+2.71m_d+0.0047/(\lambda j_d/m_d)} \nonumber \\
& & \times \left(  1-3m_d+5.2m_d^2 \right) \nonumber \\
& & \times \left( 1-0.019c_\mathrm{vir}+ 0.00025c_\mathrm{vir}^2 + 0.52/c_\mathrm{vir} \right).
\end{eqnarray}

\noindent Here, $c_\mathrm{vir}$ is the halo concentration ($c_\mathrm{vir}=r_{200}/r_{s}$, with $r_s$ being the scale radius of the halo).

N-body simulations show that the distribution of the halo spin parameter to be approximately log-normal \citep{warren92,cole96,lemson99,antonucio-delogu10}:

\begin{equation}
p(\lambda)d\lambda=\frac{1}{\sqrt{2\pi}\sigma_{\ln\lambda}}\exp\left[ -\frac{\ln^2(\lambda/ \overline{\lambda})}{2\sigma^2_{\ln\lambda}} \right] \frac{d\lambda}{\lambda}.   
\end{equation}

\noindent This means that under the assumption of a constant $j_d/m_d$ the log-normal distribution of $\lambda$ will lead to a size distribution that is approximately log-normal. 

Motivated by these theoretical considerations, Shen et al. adopt a log-normal size distribution at a given luminosity, characterized by dispersion $\sigma_{\lg\,r}$.  For faint, low-concentration systems, their results give a size dependence on r-band magnitude of $\log r_{1/2} \propto - 0.104\,M_r$ and dispersion, $\sigma_{\lg\,r} =0.19$\,dex. At the bright end, late-type galaxies have $\log r_{1/2}\propto -0.204\,M_r$ and $\sigma_{\lg\,r} =0.12$\,dex.  For comparison, these authors find that bright ($M_r<-18.5$) early-type galaxies follow $\log r_{1/2}\propto - 0.26\,M_r$ (see their equations (14), (15) and (16) for the full behavior of their functional fits).\footnote{It should be noted here that we have not attempted to convert the r-band magnitudes of \citet{shen03} into V-band magnitudes. Doing so may shift the Shen et al. relation by a small amount along the $M_V$-axis, but not so significantly that it would impact our conclusions.}  

\begin{figure}
\includegraphics[width=\hsize]{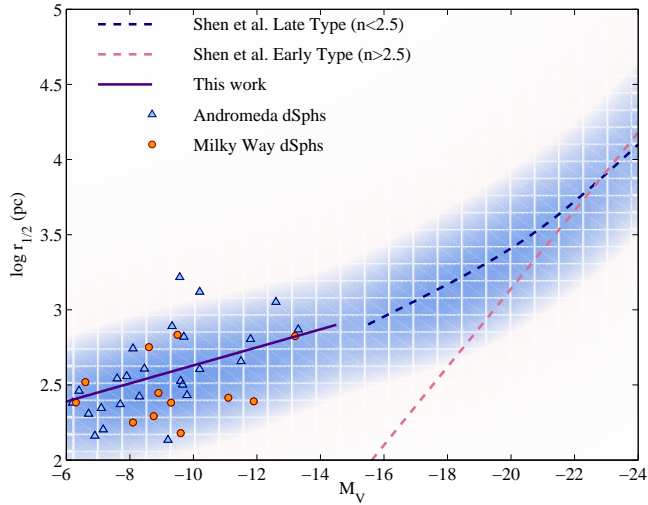}
\caption{The solid purple line shows our independent maximum likelihood fit to all Milky Way and Andromeda dSphs, while the dashed blue and dashed pink lines correspond to the \cite{shen03} relations for late-type (low-concentration) and early-type (high-concentration) galaxies, respectively. The shaded regions indicate the intrinsic scatter derived for both our work and the Shen et al. relation for late-type galaxies.  }
\label{Shen_compare}
\end{figure}

In Figure~\ref{Shen_compare} we plot our derived relation for dwarf galaxies (solid purple line), with the relations derived by  \citet{shen03} for brighter low-concentration (dashed blue line) and high-concentration (dashed pink line) galaxies in the SDSS. The shaded blue region indicates the intrinsic scatter in the derived relation for low-concentration galaxies in both our, and the Shen et al. relation.  Remarkably, the mean size, the slope, and the dispersion found in our completely independent fit to the Milky Way and Andromeda dSphs, match beautifully on to the \citet{shen03} relation for more massive low-concentration galaxies. It should be noted that the \citet{shen03} low-concentration sample actually includes dwarf elliptical galaxies (dEs) which, like dSphs, are low-concentration systems (e.g. Figure~11 of \citealt{graham08}). The size-luminosity relation we measure in the Local Group for dSphs galaxies is therefore also a continuation of that of dEs, hinting at a connection between these two types of dwarf galaxies.

\section{Simulation predictions for the sizes of faint dwarf galaxies}
\label{sec:Theoretical}

We now present theoretical predictions for the size-luminosity relation of dSphs in a $\Lambda$ Cold Dark Matter ($\Lambda$CDM) universe. These predictions are obtained by combining merger trees extracted from very high resolution N-body simulations describing the hierarchical assembly of a Milky Way-like halo, with semi analytic models (SAMs) for galaxy formation. We perform $\Lambda$CDM simulations and analyze them, including the definition of dark matter halos and the reconstruction of their detailed merger tree, using the tools described in full detail in \cite{maccio10}, hereafter M10.

\subsection{N-body code and semi analytical model}
\label{theo1}

We perform a new series of N-body simulations carried out using {\sc  pkdgrav}, a treecode written by Joachim Stadel and Thomas Quinn  \citep{stadel01}, with cosmological parameters $\Omega_{\Lambda}=0.742$, $\Omega_m=0.258$, $h=0.72$, $n=0.963$ and $\sigma_8=0.79$ \citep{komatsu09}. We select nine candidate halos from an existing low resolution dark matter simulation ($350^3$ particles within 90\,Mpc, see \citealt{neistein10}) and re-simulate them at higher resolution using the volume re-normalization technique \citep{katz93}.  The high resolution run is $12^3$ times better resolved than the low resolution one: the dark matter particle mass is $m_{d} = 3.65 \times 10^5 M_{\odot}$, where each dark matter particle has a spline gravitation (co-moving) softening of 476\,pc.

To test the importance of the host galaxies' halo mass and formation time in shaping the properties of dwarf satellites, we select three halos for each of the following dark matter masses : $M_1=0.5\times 10^{12} M_{\odot}$, $M_2=1.0\times 10^{12} M_{\odot}$, and $M_3=5.0\times 10^{12} M_{\odot}$. For each mass we then select three halos of varied concentration: one with an average value (determined using the concentration-mass relation from \citealt{maccio08} and \citealt{munoz-cuartas11}) and two with concentrations that are 1.5-$\sigma$ higher or lower, than this average. Since concentration strongly correlates with formation time (e.g. \citealt{wechsler06}) we therefore sample halos with early, average and late formation times for each mass bin. The properties of the individual halos are listed in Table \ref{table:gal}.

In order to predict the expected luminosities of satellite galaxies, we combine the results of the $N$-body simulations with the SAMs outlined in \cite{kang05,kang06} and M10.  The effects of the various physical processes in shaping the luminosity function of Milky Way satellites were studied in full detail in M10. We determined that suppression of gas infall by a photo-ionizing background, supernova feedback and tidal destruction are the most relevant processes responsible for the agreement between theoretical predictions and observational data. We characterize our best fit models as follows: (i) we regulate star formation efficiency in low-mass halos by shutting off gas cooling in structures with virial temperature below $10^4$\,K (due to the inefficiency of H$_2$ cooling); (ii) we suppress hot gas accretion in low-mass halos according to photo-ionization background, re-ionization and filtering mass arguments (\citealt{okamoto08b}); (iii) stellar feedback is modeled as a function of halo circular velocity. We keep the parameters fixed to the values providing the best fit to observations in M10\footnote{A more detailed and critical discussion of how the variation in SAM parameters may affect the final luminosity function can be found in M10.}. None of these aspects were `tuned' to match the sizes of the satellite galaxies.

\begin{figure}
\includegraphics[width=0.49\hsize]{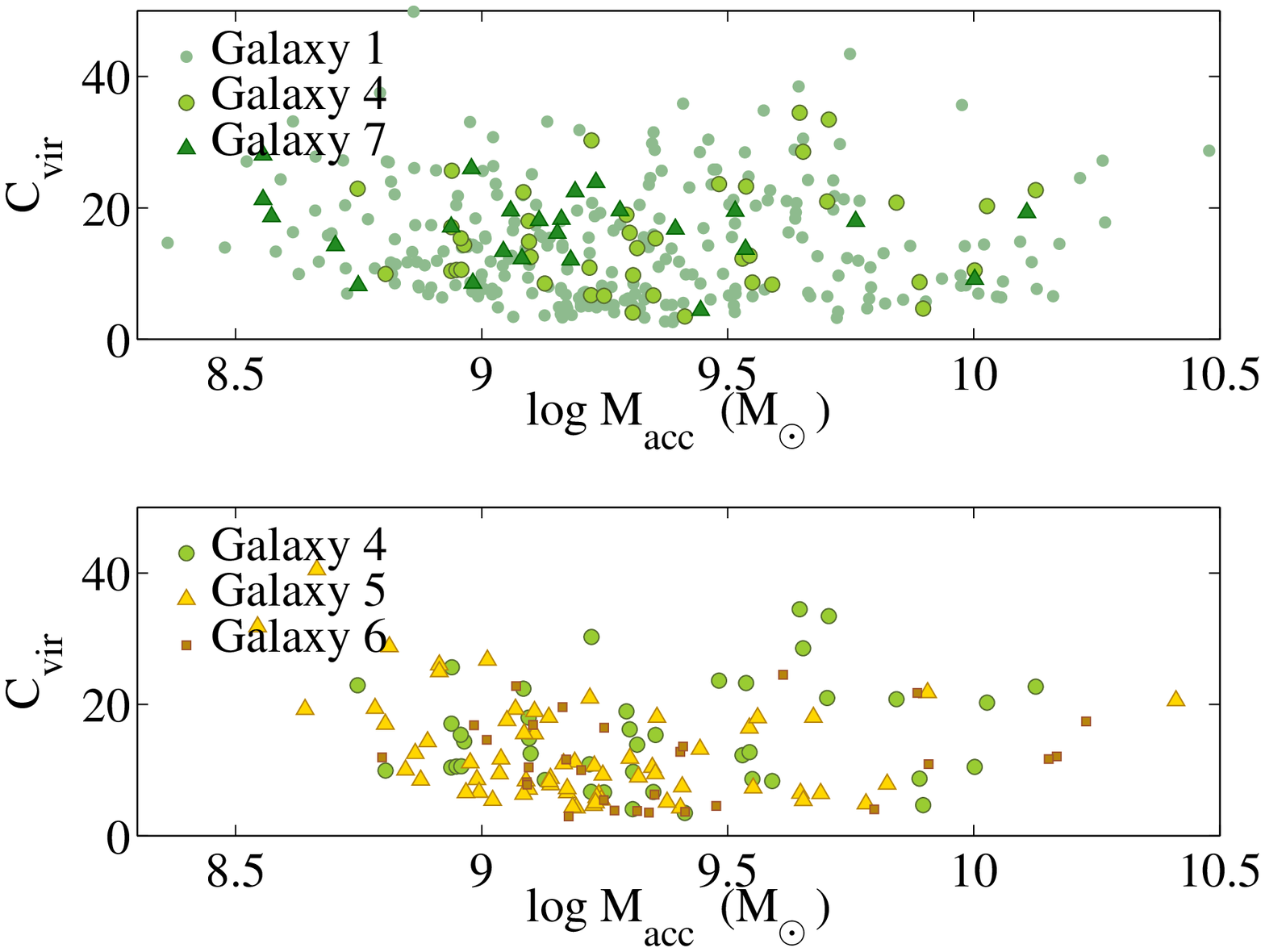}
\includegraphics[width=0.49\hsize]{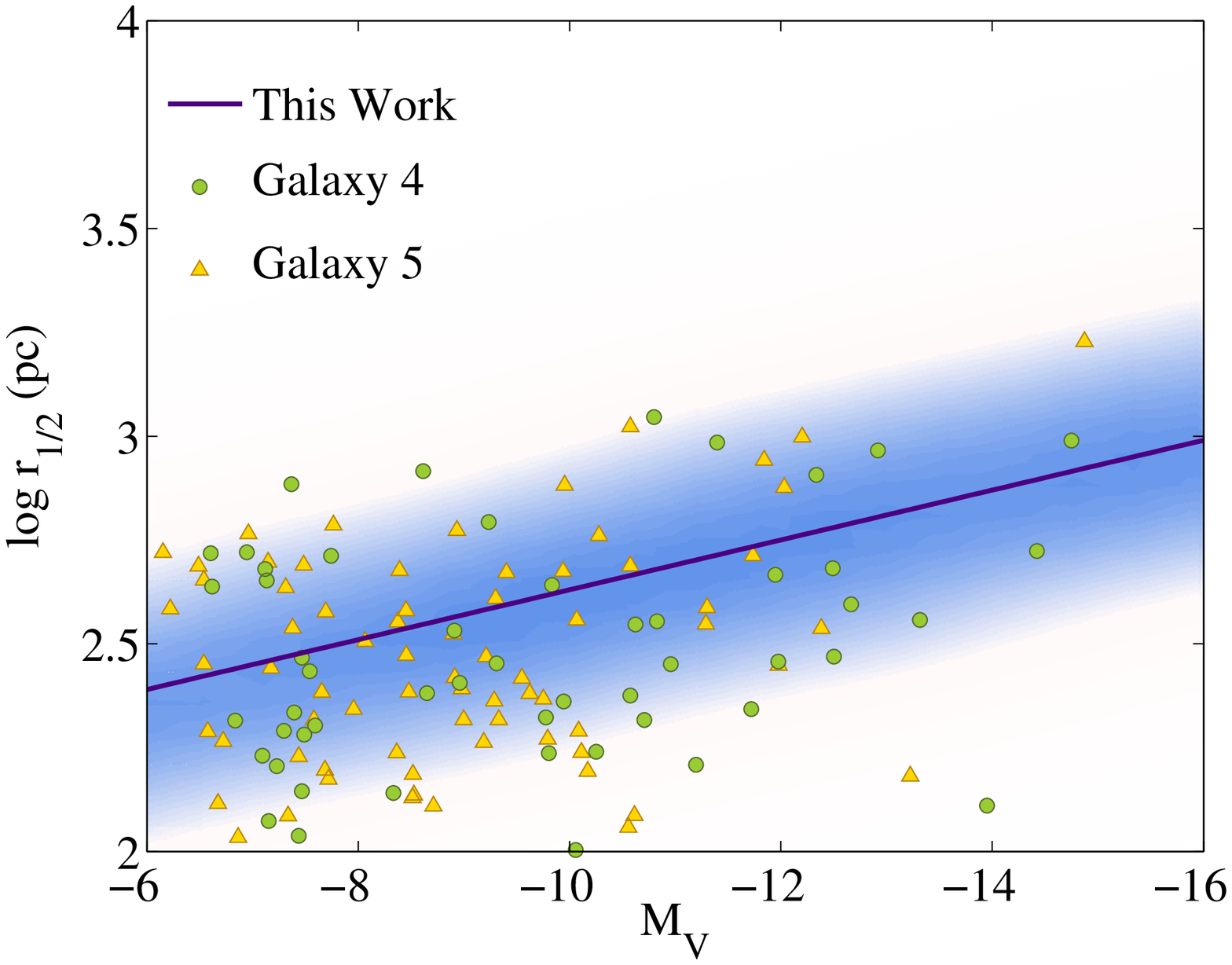}
\caption{{\bf Left:} Sub-halo dark matter mass, $M_\mathrm{acc}$, and concentration, $c_\mathrm{vir}$, at the time of accretion for a range of host halo masses and formation times (see Table \ref{table:gal} for details on each host galaxy). The upper plot shows three halos with the same formation epoch but different masses, and the lower plot shows three halos of the same mass, but different formation epochs with no noticible difference in the satellite properties. Neither change in the host-halo mass nor changes in the formation time result in a systematic difference in the concentration of the sub-halos. Note that Galaxy 1 is the highest mass, and hence shows the largest number of sub-halos. {\bf Right:} The solid purple line and the shaded region correspond to our derived relation for the combined sample of Milky Way and Andromeda dSphs. The green and yellow points correspond to the sub-halos of simulations G4 and G5. }
\label{Maccio_SAMS}
\end{figure}

We first study to what extent the mass, $M_\mathrm{acc}$, and concentration, $c_\mathrm{vir}$, of the sub-halos at the time of accretion change as a function of their host halo mass and formation time. In the upper left panel of Figure~\ref{Maccio_SAMS}, we plot the $c_\mathrm{vir}-M_\mathrm{acc}$ distributions for the sub-halo populations of three host-halos with the same formation epoch, but different masses ($0.41\times10^{12} M_{\odot}$, $0.77\times10^{12} M_{\odot}$, and $4.44\times10^{12} M_{\odot}$) and find no discernible difference between the three distributions. Similarly, we find no difference between the $c_\mathrm{vir}-M_\mathrm{acc}$ distributions of the sub-halo populations of three host-halos of equal mass but different formation epochs (see the lower-left panel of Figure~\ref{Maccio_SAMS}). 

The similarity of these $c_\mathrm{vir}-M_\mathrm{acc}$ distributions predicts indistinguishable satellite size distributions (see the dependence of $c_\mathrm{vir}$ and $M_\mathrm{acc}$ on the size of a stellar body in equation~\ref{Rd}). Therefore, within a cosmological framework we expect the same size-mass or size-luminosity relation for satellite halos orbiting host halos of different collapse times or masses\footnote{It should be noted that our simulations do not self-consistently include the presence of baryons in the main galaxy (e.g. a stellar disk), or in the satellites (a stellar core). While the presence of a stellar disk in the halo of the host galaxy can increase the mass-loss of satellites with cored matter profiles \citep{penarrubia10}, the satellite's centrally concentrated stars will make it more resilient to tidal forces \citep{schewtschenko11} and likely counteract this effect. Overall, it appears unlikely that the presence of baryons will dramatically change the results of our N-body simulations, but this should nevertheless be checked when moving beyond the simple model presented here.}.

\subsection{Comparison With Observations}

In order to compare our simulated satellites with observations of dSphs, we first need to compute the size of the stellar body and the luminosity of each sub-halo within our simulation.  While luminosity is an output of the SAM, the size of the stellar body, $R_d$, can be computed directly from N-body simulations according to equation~(\ref{Rd}). To evaluate $R_d(\lambda,c_\mathrm{vir}, m_d,  j_d,r_{200}(M_\mathrm{halo}))$ we use the values of $M_\mathrm{acc}$, $\lambda$ and $c_\mathrm{vir}$ at the time at which the satellite was accreted since in our SAM star formation is strongly suppressed after accretion (e.g. after a halo enters the virial radius of a more massive halo). Additionally, we follow \cite{shen03} in assuming that the material in a galaxy has specific angular momentum similar to that of the halo (i.e. $j_d/m_d=1$).

In the right panel of Figure~\ref{Maccio_SAMS} we show the satellite populations of two simulated galaxies (yellow triangles and green circles) overlaying our derived relation for Local Group dSphs (solid purple line, with derived intrinsic scatter shown in blue). For this comparison we have chosen two halos which are closest to the Milky Way in their properties, but, as we have shown above, there is no difference in the distributions of our simulated sub-halos so this choice is arbitrary. One can see that the simulated galaxies match the mean size and slope of our derived relation well, with only the dispersion being slightly larger, stemming from the scatter in $\lambda$. 

\section{Discussion}
\label{sec:discussion}

We set out to describe the  size-luminosity relation of Milky Way and Andromeda dSphs by a log-normal function in $r_{1/2}$, characterized by an intrinsic dispersion, $\sigma_{\lg\,r}$, and mean, $\overline{\lg\,r}$, modified to account for a slope, $S$, in magnitude. Using a maximum likelihood method which accounts for modest number statistics, observational uncertainties as well as surface brightness limitations, we determine the size-luminosity relation of the Milky Way and Andromeda dSphs both separately and as an entire population. Our empirical results can be summarized as follows:

\begin{enumerate}
\item There is no statistically significant difference between the global size-luminosity relation of Milky Way and Andromeda dSph satellites.  Indeed, the mean sizes of galaxies at a given stellar mass agree within 30\%.

Milky Way:   
\begin{equation}
\log r_{1/2}=2.38^{+0.16}_{-0.13}- 0.03\pm0.03 \times (M_V+6)
\end{equation}

Andromeda:
\begin{equation}
\log r_{1/2}=2.35^{+0.11}_{-0.14} - 0.09^{+0.02}_{-0.04} \times (M_V+6)
\end{equation}

\item We find that the size-luminosity relation obtained for dSphs is in very good agreement with the relation measured for more massive, low-concentration galaxies from \citet{shen03}.
\end{enumerate}

In a cosmological framework the first of our empirical results is comforting as our numerical simulations confirm that variations in host-halo mass or collapse-time/concentration do not lead to size differences for their sub-halos. This does not mean, however, that there do not exist Andromeda dSphs that are larger than Milky Way dSphs at a given luminosity. These exist \citep{mcconnachie06b,mcconnachie08,martin09,richardson11} but they do not translate into a significant difference in the global properties of the dSph populations. Our analysis shows that sampling, combined to the larger number of Andromeda satellites, are enough to explain the presence of larger satellites around Andromeda. One does not need to invoke different formation mechanisms, or evolution pathways, to explain apparent size differences. Hints of dynamical differences between the members of the two satellite populations \citep{collins10,kalirai10} could, however, be an tell-tale sign that the story is more complicated than displayed by their size-luminosity relation alone.

With our dSph size-luminosity relation matching so well to that derived independently for more massive late-type galaxies, our second result indicates that there may exist a common size determinant for galaxies across a wide range of masses. For disk galaxies, the size of the stellar body is generally related to the angular momentum created by torques produced during the hierarchical formation of that (satellite) halo, and the infall of material towards the center of the gravitational potential well. If a fraction of the initial angular momentum is transferred from the halo to the baryons, the baryons proceed to cool into a disk until they are prevented from further collapsing by the angular momentum of the material. This results in a relation between the final size of the galaxy and stellar mass, which is itself related to its dark matter halo mass and the initial angular momentum of the baryons.

\citet{shen03} were able to explain the size-luminosity relation of massive late-type ($n<2.5$) galaxies by applying the angular momentum formulation of \citet{mo98}. Dwarf spheroidal galaxies, like dwarf elliptical galaxies, show S\'ersic indices of $n \sim1$ (e.g. \citealt{derijcke03}), making them a faint extension of late-type systems by this criterion. When we apply this simple angular momentum formulation derived for more massive disk galaxies to the scale of dSphs, we are able to reproduce the observed size-luminosity distribution of dSph galaxies in both mean size, slope and scatter (see Figure~\ref{Maccio_SAMS}). This strongly suggests that angular momentum arguments and the cosmological framework play an important role in setting the sizes of dSphs.

At face value, such an explanation for the sizes of dSphs implies that they should be, or have been, rotating. While rotation is not manifest in the existing data for many galaxies, there are signs of significant net angular momentum in emergent data sets. \cite{battaglia08} has found a radial velocity gradient in the Sculptor dwarf galaxy, which they interpret as an indication of internal rotation, and \citet{fabrizio11} find the radial velocity distribution across the body of the Carina dSph to show a radial gradient that they take as possible evidence of rotation. The explanation for these radial velocity gradients is still debated, though, as they could simply be the result of the dwarf galaxies' proper motions \citep{kaplinghat08,walker08,strigari10}.

Even if more subtle signs of rotation could be discovered in the future, the question arises of how clear signatures of rotation should be. The flatness and coldness of a gas disk, and the resulting stellar disk, is given by the ratio of its rotation velocity to its velocity dispersion ($v_{rot}/\sigma$). Clearly, for low mass galaxies, this ratio will approach unity if the rotation velocities approach the turbulent velocities of the ISM, $\sim 5\kms$. In addition, much of the initial rotation in dwarf galaxies may have been erased if they have been tidally stirred via their interactions with their host galaxy. The recent simulations of \citet{lokas11} postulate that such stirring indeed could or should be effective in erasing the rotational signatures in dSphs while, at the same time, only mildly reducing the size of their stellar component. Such a conclusion also follows the work of \cite{mayer01b,mayer01a} who initially proposed that dwarf irregular galaxies are actually the progenitors of dSphs, transformed through tidal stirring. This has some observational evidence, as it explains the fact that dwarf irregulars are found on the outskirts of their host galaxies, while dSphs are mainly located further in. 

The scenario we propose for the sizes of low-mass satellite galaxies would therefore start out with a well founded theory based on angular momentum considerations: dSph galaxies form as disks with sizes set by the angular momentum, and initially have $v_{rot}/\sigma \simeq1$. Tidal stirring would then randomize some fraction of their kinematics, leaving their disks with little sign of rotation, which is consistent with the data. In this theory, more distant satellites, which feel weaker tides should, on average, show more rotation, as seems to be the case with the Cetus dwarf galaxy, an isolated Local Group dSph that shows a rotation signature of $8\kms$ \citep{lewis07}, consistent with this scenario.

\section{Acknowledgments}
We thank Glenn van de Ven for helpful discussions and Alan McConnachie for commenting on this work. We also wish to acknowledge the important role played by the PAndAS collaboration in significantly expanding our knowledge of the M31 satellite system. All numerical simulations have been performed on the THEO cluster of the Max-Planck-Institut f\"ur Astronomie at the Rechenzentrum in Garching. This work was supported by the National Science and Engineering Research Council of Canada. NFM acknowledges funding by Sonderforschungsbereich SFB 881 ``The Milky Way System'' (subproject A3) of the German Research Foundation 
(DFG).



\clearpage
\begin{deluxetable}{l l l l}
\tablecaption{Properties of Milky Way dSphs  \label{mw_table}}
\tablewidth{0pt}
\tablehead{ \colhead{dSph} & \colhead{$M_V$} & \colhead{$r_{1/2}$ (pc)\tablenotemark{1}}& \colhead{Reference} }
\startdata
Bo\"otes~I			&$-6.3\pm0.2$		&	$242^{+22}_{-20}$ 	&\cite{martin08b}\\ 
Bo\"otes~II		&$-2.7\pm0.9$		&	$51\pm17$	 	&\cite{martin08b}\\
Carina 			&$-9.3$			&	$241\pm23$	 	&\cite{irwin95}, \cite{mateo98}\\
Canes Venatici~I	&$-8.6^{+0.2}_{-0.1}$	&	$564\pm36$	 	&\cite{martin08b}\\
Canes Venatici~II	&$-4.9\pm0.5$		& 	$74^{+14}_{-10}$ 	&\cite{martin08b}\\
Coma Bernices		&$-4.1\pm0.5$		& 	$77\pm10$	 	&\cite{martin08b}\\
Draco			&$-8.75\pm0.05$	&	$196\pm12$	 	& \cite{martin08b}\\
Fornax			&$-13.2$			&	$668\pm34$ 	 	&\cite{irwin95}, \cite{mateo98}\\
Hercules		 	&$-6.6\pm0.3$		&	$330^{+75}_{-52}$ 	&\cite{martin08b}\\
Leo~I			&$-11.9$			&	$246\pm19$   	 	&\cite{irwin95}, \cite{mateo98}\\
Leo~II			&$-9.6$			&	$151\pm17$   	 	&\cite{irwin95}, \cite{mateo98}\\
Leo~IV			&$-5.8\pm0.4$		&	$206\pm36$ 		&\cite{dejong10a}\\
Leo~V			&$-5.2\pm0.4$		& 	$133\pm31$		&\cite{dejong10a}\\
Leo~T			&$-8.1\pm0.1$		&	$178\pm39$	 	&\cite{martin08b}\\
Sculptor			&$-11.1$			&	$260\pm39$   	 	&\cite{irwin95}, \cite{mateo98}\\  
Segue~1			&$-1.5^{+0.6}_{-0.8}$	&	$29^{+8}_{-5}$	 	&\cite{martin08b}\\
Segue~2			&$-2.5\pm0.3$		&	$34\pm5$		 	&\cite{belokurov09}\\
Sextans			&$-9.5$			&	$682\pm117$ 	 	&\cite{irwin95}, \cite{mateo98}\\
Ursa Major~I		&$-5.5\pm0.3$		& 	$318^{+50}_{-39}$ 	&\cite{martin08b}\\	
Ursa Major~II		&$-4.2\pm0.5$		& 	$140\pm25$    	  	&\cite{martin08b}\\
Ursa Minor		&$-8.8$			&	$280\pm15$   	 	&\cite{irwin95}, \cite{mateo98}\\
\enddata
\tablenotetext{1}{Where $r_{1/2}$ is the half-light radius measured along the semi-major axis.}
\end{deluxetable}

\begin{deluxetable}{l l l l}
\tablecaption{Properties of Andromeda dSphs \label{and_table}}
\tablewidth{0pt}
\tablehead{ \colhead{dSph} & \colhead{M$_V$} & \colhead{$r_{1/2}$(pc)\tablenotemark{1}} &\colhead{Reference}}
\startdata
Andromeda~I		&$-11.8\pm0.1$		&	$638^{+34}_{-17}$			&\cite{mcconnachie06b}\\
Andromeda~II		&$-12.6\pm0.2$		&	$1126\pm17$				&\cite{mcconnachie06b}\\
Andromeda~III		&$-10.2\pm0.3$		&	$403\pm17$				&\cite{mcconnachie06b}\\
Andromeda~V		&$-9.6\pm0.3$		&	$336\pm17$				&\cite{mcconnachie06b}\\
Andromeda~VI		&$-11.5\pm0.2$		&	$454\pm17$				&\cite{mcconnachie06b}\\
Andromeda~VII		&$-13.3\pm0.3$		&	$739\pm34$				&\cite{mcconnachie06b}\\
Andromeda~IX		&$-8.1^{+0.4}_{-0.1}$	&	$552^{+22}_{-110}$			&\cite{collins10}	\\
Andromeda~X		&$-7.36\pm0.07$	&	$235\pm20$				&\citet{brasseur11a}\\
Andromeda~XI		&$-6.9^{+0.5}_{-0.1}$	&	$145^{+24}_{-20}$			&\cite{collins10}	\\
Andromeda~XII		&$-6.4^{+0.1}_{-0.5}$	&	$289^{+70}_{-47}$			&\cite{collins10}	\\
Andromeda~XIII	&$-6.7^{+0.4}_{-0.1}$&	$203^{+27}_{-44}$			&\cite{collins10}	\\
Andromeda~XIV	&$-8.46\pm0.07$	&	$406\pm27$				&\cite{brasseur11b}\\
Andromeda~XV		&$-9.8\pm0.4$		&	$270\pm30$				&\cite{ibata07}, \cite{letarte09}\\
Andromeda~XVI	&$-9.2\pm0.5$		&	$136\pm14$				&\cite{ibata07}, \cite{letarte09}\\
Andromeda~XVII	&$-8.30\pm0.06$	&	$265\pm19$				&\citet{brasseur11a}\\
Andromeda~XVIII	&$-9.66\pm0.07$	&	$317\pm23$				&\cite{brasseur11b}\\
Andromeda~XIX	&$-9.57\pm0.69$	&	$1647\pm252$				&\cite{brasseur11b}\\
Andromeda~XX		&$-7.16\pm0.10$	&	$160\pm30$				&\cite{brasseur11b}\\
Andromeda~XXI	&$-9.33\pm0.07$	&	$777\pm53$				&\cite{brasseur11b}\\
Andromeda~XXII	&$-6.2\pm0.15$		&	$405\pm27$				&\cite{brasseur11b}\\
Andromeda~XXIII	&$-10.2\pm0.5$		&	$1316\pm75$\tablenotemark{2}	&\cite{richardson11}\\
Andromeda~XXIV	&$-7.6\pm0.5$		&	$349\pm19$\tablenotemark{2}	&\cite{richardson11}\\
Andromeda~XXV	&$-9.7\pm0.5$		&	$661\pm37$\tablenotemark{2}	&\cite{richardson11}\\
Andromeda~XXVI	&$-7.1\pm0.5$		&	$222\pm12$\tablenotemark{2}	&\cite{richardson11}\\
Andromeda~XXVII	&$-7.9\pm0.5$		&	$361\pm21$\tablenotemark{2}	&\cite{richardson11}\\
\enddata
\tablenotetext{1}{Where $r_{1/2}$ is the half-light radius measured along the semi-major axis.}
\tablenotetext{2}{The uncertainties on these values are based on the distance uncertainties listed in \citet{richardson11}.}
\end{deluxetable}

\begin{table}
\caption{Simulated Galaxy Parameters}
\begin{tabular}{lcrccc}
\hline  Halo &  Mass ($10^{12}M_{\odot}$) & $N_\mathrm{particles}$ & $R_\mathrm{vir}$ (kpc) & Formation time  \\
\hline
G1 & 4.44 & 12,157,818 & 339.5 & average  \\
G2 & 4.46 & 12,103,298 & 339.9 & high  \\
G3 & 4.89 & 13,270,207 & 349.8 & low  \\
G4 & 0.77 &  1,889,570 & 189.8 & average  \\
G5 & 0.89 &  2,184,048 & 199.2 & high  \\
G6 & 0.81 &  1,987,729 & 192.2 & low  \\
G7 & 0.41 &  1,043,557 & 154.4 & average  \\
G8 & 0.48 &  1,221,726 & 161.3 & high  \\
G9 & 0.51 &  1,298,083 & 165.0 & low  \\
\hline
\label{table:gal}
\end{tabular}
\end{table}




\end{document}